\documentclass[pre,10pt, twocolumn, aps, superscriptaddress, showpacs, groupaddress]{revtex4-2}
\usepackage[T1]{fontenc}
\usepackage{bigints}
\usepackage{amssymb}
\usepackage{graphicx}
\usepackage{amsmath}
\usepackage{dcolumn}
\usepackage{multirow}
\usepackage{hyperref}
\usepackage[normalem]{ulem}
\usepackage{physics}
\usepackage{bm}
\usepackage{comment}
\usepackage{xcolor}
\usepackage[caption=false]{subfig}
\usepackage{siunitx}
\usepackage{dsfont}
\usepackage{bbold}
\usepackage[toc]{appendix}
\usepackage{relsize}
\usepackage{leftidx}
\usepackage{bbm}
\usepackage{stackengine}
\usepackage{float}
\usepackage{tikz}
\usetikzlibrary{arrows.meta,calc,decorations.pathmorphing,positioning,fit,backgrounds}

\definecolor{lapislazuli}{rgb}{0.15, 0.38, 0.61}
\definecolor{YKblue}{rgb}{0.0, 0.18, 0.65}
\definecolor{carmine}{rgb}{0.81, 0.09, 0.13}
\definecolor{lavender}{rgb}{0.84, 0.79, 0.87}

\hypersetup{
	colorlinks=true,
	linkcolor=blue,
	linktoc=page,
	citecolor=blue,
	urlcolor=blue
}

\begin{document}

\title{Axion magnetohydrodynamics and reconnection‑driven axion bursts}

\author{Hugo Ter\c{c}as}
\affiliation{Instituto Superior de Engenharia de Lisboa,
  Instituto Polit\'{e}cnico de Lisboa,
  Rua Conselheiro Em\'{i}dio Navarro, 1959-007 Lisboa, Portugal}
\affiliation{GoLP/Instituto de Plasmas e Fus\~{a}o Nuclear,
  Instituto Superior T\'{e}cnico, Universidade de Lisboa,
  1049-001 Lisboa, Portugal}

\begin{abstract}
We formulate axion magnetohydrodynamics beyond the ideal limit, retaining axion inertia and the essential physics of non--ideal plasmas from first principles. In this framework, regions where magnetic flux freezing breaks down acquire a new physical role: whenever $\mathbf{E}\!\cdot\!\mathbf{B}\neq0$, magnetic dissipation acts as a localized source of axion radiation. We show that magnetic reconnection naturally excites mixed Alfv\'en--axion modes, enabling coherent energy exchange between magnetic fields and axions in magnetically dominated environments. In neutron stars and magnetars, this mechanism leads generically to transient axion bursts powered by reconnection--driven Alfv\'enic dissipation. We connect this production process to observational prospects and derive a characteristic sensitivity to the axion--photon coupling, complementary to searches based on static magnetic fields.
\end{abstract}

\maketitle

\textit{Introduction.---}
The axion occupies a singular position in contemporary fundamental physics.
Originally conceived as an elegant resolution of the strong CP problem in quantum
chromodynamics, it has since emerged as one of the most compelling and economical
candidates for the dark matter of the Universe
\cite{PecceiQuinn1977a,PecceiQuinn1977b,Weinberg1978,Wilczek1978,
PreskillWiseWilczek1983,AbbottSikivie1983,DineFischler1983,Visinelli2009, Marsh2016,DiLuzio2020}.
Its weak but generic coupling to electromagnetism has motivated an extensive program
of astrophysical searches, in which large--scale magnetic fields act as natural
amplifiers of axion signatures through axion--photon conversion
\cite{Raffelt1996,IrastorzaRedondo2018,Graham2015,Witte2020}.
In this context, compact objects endowed with strong magnetic fields have long been
recognized as particularly promising axion laboratories, enabling both the
production of axions and the conversion between axions and photons \cite{Hook2018, Roy2026}.

Within this broad landscape, neutron stars and magnetars stand out as extreme and
unique environments.
With surface magnetic fields reaching $10^{14}$--$10^{15}\,\mathrm{G}$ and hosting
dense, relativistic magnetospheres, magnetars have been extensively explored as
sites for axion dark matter conversion, resonant axion--photon mixing in plasma, and
axion--mediated radio emission
\cite{Prabhu2021,Buckley2021, Prabhu2023, Caputo2024}. Nevertheless, in most existing studies the electromagnetic fields are prescribed
externally and the axion is treated as a passive degree of freedom, converting or
propagating in a fixed plasma background without participating dynamically in the
energy budget of the system \cite{Millar2021, Tjemsland2024}.

A complementary line of research has begun to relax this assumption by considering
the backreaction of axions on plasma dynamics.
A growing body of work has shown that axions can modify wave propagation,
instabilities, and collective plasma modes through axion electrodynamics
\cite{Huang2015,Xia2016,FluggeZhukov2018,TercasPRL2018,
MendoncaPRD2020, HwangNoh2022}.
Yet much of this literature remains anchored to idealized or adiabatic limits, in
which dissipative plasma processes are suppressed and genuine energy exchange
between magnetic fields, plasmas, and axions is excluded by construction.
Magnetars, by contrast, are characterized by violent and recurrent activity in the
form of bursts and giant flares \cite{Sathyaprakash2024, Yakovlev2024, Pacholski2026}, during which vast reservoirs of magnetic energy are
released on dynamical timescales
\cite{DuncanThompson1992,ThompsonDuncan1995,Turolla2015,KaspiBeloborodov2017, Beloborodov2017}. A central role in this activity is played by magnetic reconnection, an intrinsically
non--ideal plasma process in which magnetic field lines change topology and magnetic
energy is rapidly converted into particle acceleration, heating, and radiation
\cite{Parker1957,Sweet1958,Petschek1964,Biskamp1986,ZweibelYamada2009}.
Crucially, reconnection is characterized by localized regions where magnetic flux
freezing breaks down and the condition $\mathbf{E}\!\cdot\!\mathbf{B}\neq0$ is
unavoidably satisfied.
In the presence of a dynamical axion field, such regions act as localized sources in
the axion field equation, allowing magnetic energy to be converted directly into
propagating axion excitations.
Magnetic reconnection is thus elevated from a purely plasma--physical process to a
potential engine of axion production in extreme astrophysical environments.

In this Letter, we formulate axion magnetohydrodynamics (aMHD) beyond the ideal
limit, retaining axion inertia and the essential microphysics of non--ideal plasmas
from first principles.
We show that magnetic reconnection in magnetar magnetospheres provides a natural and
efficient channel for axion production through the excitation and dissipation of
Alfv\'enic disturbances, complementing existing scenarios based on axion conversion
in prescribed magnetic backgrounds.
This framework clarifies the limits of previous idealized approaches and reveals a
regime in which plasma dynamics, magnetic topology, and axion physics become
intrinsically entwined, with direct implications for magnetars, axion searches, and
the physics of extreme plasmas.

\textit{Axion Magnetohydrodynamics.---} We begin by formulating axion magnetohydrodynamics (aMHD) by retaining the ingredients that are essential for dissipation, energy transfer, and axion production. We consider a relativistic axion field $a(x)$, with $x=(t,\mathbf{x})$ ($c=1$), coupled to electromagnetism through the standard axion--photon interaction.
The dynamics is governed by the Lagrangian 
\begin{align}
\mathcal{L} &= 
-\frac{1}{4} F_{\mu\nu}F^{\mu\nu}
+ \frac{1}{2}\partial_\mu a\,\partial^\mu a
- \frac{1}{2} m_a^2 a^2 + \frac{g_{a\gamma}}{4}\, a\, F_{\mu\nu}\tilde F^{\mu\nu}\nonumber \\[5pt]
& + \mathcal{L}_{\mathrm{plasma}},
\label{eq:action}
\end{align}
where $F_{\mu\nu}=\partial_\mu A_\nu-\partial_\nu A_\mu$ is the electromagnetic field tensor, $\tilde F^{\mu\nu}$ its dual, $g_{a\gamma}$ the axion--photon coupling constant, and $\mathcal{L}_{\mathrm{plasma}}$ describes the charged plasma constituents. Variation of the action with respect to the electromagnetic four--potential yields the modified Maxwell equations,
\begin{align}
&\bm\nabla\!\cdot\!\left(\mathbf{E}+g_{a\gamma}\mathbf{B}\cdot \bm\nabla a\right)=\frac{\rho}{\epsilon_0}, \\[10pt]
&\bm\nabla\!\times\!\mathbf{B}-\partial_t\mathbf{E}
=\mu_0\mathbf{J}
+g_{a\gamma}\!\left(\dot a\,\mathbf{B}+\bm\nabla a\times\mathbf{E}\right),
\end{align}
where $\rho$ and $\mathbf{J}$ denote the charge and current densities of the plasma. The axion thus induces effective charge and current densities proportional to gradients and time derivatives of $a$, reflecting the parity--odd structure of the interaction.

Variation with respect to the axion field yields a massive Klein--Gordon equation with a topological source,
\begin{equation}
\left(\Box+m_a^2\right)a
=-\,g_{a\gamma}\,\mathbf{E}\!\cdot\!\mathbf{B},
\label{eq:axion_eom}
\end{equation}
with $\Box=\partial_t^2-\nabla^2$ the d'Alembert operator.
Equation~\eqref{eq:axion_eom} lies at the core of the present work: regions where $\mathbf{E}\!\cdot\!\mathbf{B}\neq0$ act as localized sources of axion excitations, enabling the direct conversion of electromagnetic energy into propagating axion radiation.
Crucially, the second--order time derivative encodes axion inertia and cannot be neglected whenever dynamics occur on timescales shorter than $m_a^{-1}$. To close the system, the plasma dynamics must be specified.
While a fully kinetic description is possible, it suffices here to adopt a two--fluid model for electrons and ions.
Combining the corresponding equations of motion yields a generalized Ohm’s law incorporating resistivity, Hall physics, pressure gradients, and electron inertia \cite{Braginskii,Nicholson,Fitzpatrick},
\begin{equation}
\mathbf{E}+\mathbf{v}\times\mathbf{B}
=\eta_{\rm el}\,\mathbf{J}
-\frac{\mathbf{J}\times\mathbf{B}}{ne}
-\frac{\bm\nabla p_e}{ne}
-\frac{m_e}{e} D_t^e\mathbf{v}_e,
\label{eq:Ohm_general}
\end{equation}
with $D_t^e=\partial_t +{\bf v}_e \cdot \bm \nabla$ denoting the Lagrange derivative and $\eta_{\rm el}$ being an effective resistivity encoding non--ideal plasma effects \cite{Nicholson, Fitzpatrick}. In magnetically dominated environments such as magnetar magnetospheres, the relevant dynamics occurs on scales $L\gg c/\omega_{pi}$ and timescales $\omega\ll\omega_{pe}$, while the plasma is strongly low--$\beta$.
Under these controlled conditions, the Hall term, pressure gradients, and electron inertia may be consistently neglected, and the generalized Ohm’s law reduces to
\begin{equation}
\mathbf{E}+\mathbf{v}\times\mathbf{B}=\eta_{\rm el}\,\mathbf{J}.
\label{eq:Ohm}
\end{equation}
Axion magnetohydrodynamics emerges when Eqs.~\eqref{eq:axion_eom} and~\eqref{eq:Ohm} are combined with Maxwell’s equations.
Neglecting displacement current effects in Amp\`ere’s law and retaining terms linear in $g_{a\gamma}$, the resulting dynamical equations read
\begin{align}
&\frac{\partial \mathbf{B}}{\partial t}
=\bm\nabla\times(\mathbf{v}\times\mathbf{B})
-\eta\,\bm\nabla\times(\bm\nabla\times\mathbf{B})
+\eta g_{a\gamma}\,\bm\nabla\!\left(\dot a\,\mathbf{B}\right), \nonumber\\[5pt]
&\left(\Box+m_a^2\right)a
=-\,g_{a\gamma}\eta\,(\bm\nabla\times\mathbf{B})\!\cdot\!\mathbf{B},
\label{eq:MHD_fields}
\end{align}
where $\eta=\eta_{\rm el}/\mu_0$ is the magnetic diffusivity. The magnetic and axion fields are coupled to the plasma through the fluid conservation laws,
\begin{align}
&\frac{\partial \rho}{\partial t} + \bm \nabla\cdot(\rho\,\mathbf{v}) =0,\nonumber \\[5 pt]
&(\rho+w) \left(\frac{\partial \mathbf{v}}{\partial t} +
\mathbf{v}\cdot \bm\nabla\mathbf{v} \right) = -\bm \nabla p + \mathbf{J}\times\mathbf{B},
\label{eq:MHD_fluid}
\end{align}
where $\rho$ is the rest--mass density, $p$ the plasma pressure, and $w = B^2+(\gamma -1)p/\gamma$ the relativistic enthalpy density. Equations \eqref{eq:MHD_fields} and \eqref{eq:MHD_fluid} constitute the central result of this paper. In the ideal limit $\eta\to 0$, axions decouple from the plasma at large scales; conversely, whenever non--ideal electric fields develop, helicoidal magnetic structures act as unavoidable sources of axion radiation.

Two important comments are in order. First, we contrast this formulation with earlier treatments of axion--modified magnetohydrodynamics, notably in cosmological and relativistic MHD contexts \cite{HwangNoh2022}.
In such approaches, the axion field is commonly assumed to evolve adiabatically and is treated as a coherent background. The axion inertia term is then adiabatically eliminated, rendering the theory insensitive to rapid, localized plasma processes. Second, the resistivity $\eta$ appearing in Eq.~\eqref{eq:Ohm} is not that proposed by Spitzer, $\eta_{\rm Spitzer}\sim m_e\nu_{ei}/n_e e^2\propto T_e^{-3/2}$ \cite{Spitzer1962}, since in the environments of interest dissipation arises from kinetic processes such as
current--driven instabilities, electron inertia, and collisionless reconnection,
which generate localized parallel electric fields on microscopic scales
\cite{PriestForbes2000, Uzdensky2011}. These effects may be consistently coarse--grained into an effective macroscopic
resistivity, allowing the large--scale dynamics to be described within a controlled
MHD framework while preserving the essential physics of axion production.

\textit{Axion$-$ Alfvén \& Magnetosonic waves.}  We consider a homogeneous equilibrium with constant density $\rho_0$ and pressure $p_0$, a uniform background magnetic field $\mathbf{B}_0$, and a plasma initially
at rest. The background axion field is taken to be homogeneous, $\nabla a_0=0$, with slow
temporal evolution. First-order perturbations are introduced in Eqs. \eqref{eq:MHD_fields} and \eqref{eq:MHD_fluid} as
$\rho = \rho_0 + \delta\rho$, $\mathbf{v} = \delta\mathbf{v}$, $\mathbf{B} = \mathbf{B}_0 + \delta\mathbf{B}$, and $a = a_0 + \delta a$, with
and all perturbed quantities are assumed to vary as plane waves $\exp[i(\mathbf{k}\cdot\mathbf{x}-\omega t)]$. 
Interestingly, the axion couples {\it selectively} to magnetic perturbations carrying finite current helicity, providing a new channel for wave hybridization in non--ideal plasmas, as we show below.

Shear Alfv\'en waves correspond to transverse velocity and magnetic perturbations polarised perpendicular to both the wavevector and the background magnetic field.
They play a central role in magnetically dominated plasmas and, in particular, in reconnection--driven dynamics. We first focus on the case of propagation parallel to the background magnetic field,
$\mathbf{k}\parallel\mathbf{B}_0$, for which compressive fluctuations decouple identically. The linear dynamics then reduces to the subspace spanned by the transverse
velocity component and the axion perturbation. The resulting dispersion relation reads
\begin{equation}
(\omega^2-k^2 v_A^2+i\eta k^2\omega)
(\omega^2-k^2-m_a^2)-
g_{a\gamma}^2\eta^2 B_0^2 k^2\omega^2
=0 ,
\label{eq:alfven_full}
\end{equation}
where
\(
v_A = B_0/\sqrt{B_0^2+w_0}
\)
is the relativistic Alfv\'en speed. In the absence of axions, Eq.~\eqref{eq:alfven_full} factorises into the standard resistive Alfv\'en mode and a free massive axion branch. For finite axion--photon coupling, the two branches hybridise through dissipative,
helicity--selective terms proportional to $g_{a\gamma}\sqrt{\eta}$.
The hybridisation is strongest when the Alfv\'en frequency,
$\omega\simeq k v_A$, approaches the axion mass gap $m_a$. The strength of the coupling is naturally quantified by the
Alfv\'en--axion Rabi frequency, $\Omega_A=
g_{a\gamma}\,\eta\,v_A \,B_0 \,m_a $, for which a fiducial estimate reads
\begin{align}
\Omega_A
\simeq
4\,\mathrm{mHz}\;
\left(\frac{g_{a\gamma}}{10^{-12}\,\mathrm{GeV^{-1}}}\right)
\left(\frac{B_0}{10^{15}\,\mathrm{G}}\right)
\nonumber\\ \left(\frac{\eta }{10^{-2}\,\mathrm{m^2 ~s^{-1}}}\right)
\left(\frac{v_A}{1}\right)
\left(\frac{m_a}{10^{-6}\,\mathrm{eV}}\right),
\end{align}   
which measures the rate of coherent energy exchange between magnetic and axionic
degrees of freedom.
The corresponding eigenmodes are mixed Alfv\'en--axion polaritons.
Importantly, the coupling vanishes continuously in the ideal MHD limit
$\eta\to 0$, highlighting the crucial role of non--ideal electric fields.

To consider compressive magnetosonic perturbations, the wavevector is taken to form an angle $\theta$ with the background magnetic field, $\mathbf{k} = k(\cos\theta, 0, \sin\theta)\equiv(k_\perp,0,k_\parallel)$. In contrast to shear Alfv\'en waves, magnetosonic modes involve compressive motions and therefore couple longitudinal and transverse velocity fluctuations, $(\mathbf{k}\times\delta\mathbf{B})\cdot\mathbf{B}_0
\;\propto\;
k_\parallel\,k_\perp\,\delta B$, and thus vanishes for purely parallel $(\theta=0)$ or purely perpendicular $(\theta=\pi/2)$
propagation. Magnetosonic--axion coupling is therefore intrinsically oblique.
\begin{equation}
\mathcal{D}_{\rm MS}(\omega,k)\,
(\omega^2-k^2-m_a^2)
-
\Omega_{\rm MS}^2\,\omega^2 k^2 
=0, 
\label{eq:disp_ms_axion}
\end{equation}
where $\mathcal{D}_{\rm MS}(\omega, k)
\equiv \omega^4 - \omega^2 k^2\!\left(v_A^2 + c_s^2\right)
+ k^4 v_A^2 c_s^2 \cos^2\theta =0$ defines the bare magnetosonic dispersion. The strength of the coupling is controlled by the Alfv\'en--axion mixing scale $\Omega_{\rm MS}
= \Omega_A\sin\theta\cos \theta$. In magnetically dominated plasmas, where $v_A\simeq c_s\simeq1$, the fast magnetosonic branch carries a substantial transverse magnetic component and is therefore the mode most
efficiently coupled to the axion field.
By contrast, the slow magnetosonic mode, which becomes predominantly field--aligned in this
regime, is only weakly affected.
As in the Alfv\'en case, magnetosonic--axion hybridisation vanishes continuously in the ideal
MHD limit $\eta\to0$, underscoring the essential role of non--ideal electric fields in axion
production.

\textit{Axion production from magnetic reconnection.---} The mechanism discussed here does not rely on the slow Sweet--Parker reconnection
regime, which is controlled by classical resistive diffusion and leads to
vanishingly small reconnection rates in high--Lundquist--number plasmas
\cite{Sweet1958,Parker1963,Biskamp1986}.
Instead, magnetar magnetospheres are widely believed to operate in fast reconnection
regimes, such as plasmoid--dominated or turbulent reconnection, in which magnetic
energy is released on Alfv\'enic timescales and efficiently channelled into
propagating Alfv\'enic disturbances
\cite{Uzdensky2010, LouHuang2013,Bhattacharjee2009,Lazarian1999, Loureiro2012}.
It is precisely this rapid, wave--mediated form of reconnection that provides the
helicity, dissipation, and dynamical timescales required for efficient axion
production within the axion magnetohydrodynamic framework developed here
\cite{ThompsonDuncan2001,Lyutikov2003,Lyutikov2006,BransgroveBeloborodov2026}. Magnetic reconnection is intrinsically a non--ideal plasma process, characterized by
localized regions in which the ideal MHD condition
$\mathbf{E}+\mathbf{v}\times\mathbf{B}=0$ breaks down and parallel electric fields
$\mathbf{E}\!\cdot\!\mathbf{B}\neq0$ are generated.
In strongly magnetized, relativistic plasmas such as magnetar magnetospheres,
reconnection is widely understood to proceed through the excitation, propagation,
and dissipation of Alfv\'{e}n waves, which transport magnetic helicity and
field--aligned currents away from reconnecting current sheets
\cite{ThompsonDuncan1995,ThompsonDuncan2001,Lyutikov2003,Lyutikov2006,
Uzdensky2011,PontinPriest2022,BransgroveBeloborodov2026}.
Within the aMHD framework derived above, this non--ideal
dynamics acts as a source of axions via the dissipative helicity term $-g_{a\gamma}\eta (\bm\nabla\times {\bf B})\cdot {\bf B}.$ Axion production is therefore not a secondary effect imposed on a prescribed
electromagnetic background, but a direct consequence of magnetic dissipation. To make this connection explicit, we consider an Alfv\'{e}n wave generated in
a reconnecting current sheet, propagating along to the background magnetic field, ${\bf k}\parallel \mathbf{B}_0$. The transverse magnetic perturbation may be written as $\mathbf{B}_\perp(\mathbf{x},t)=
\Re\!\left[
\delta B_A
(\hat{\mathbf{x}}\pm i\hat{\mathbf{y}})
e^{i(kz-\omega_A t)}
\right]$, yielding
$(\bm\nabla\times\mathbf{B})\!\cdot\!\mathbf{B}=
k B_0\,\delta B_A,$ and demonstrating that Alfv\'{e}n waves generated by reconnection carry finite current
helicity and therefore provide efficient axion sources. Substituting in Eq.~\eqref{eq:MHD_fields}, we obtain
$(\omega^2-k^2-m_a^2)\,a_k =-\,g_{a\gamma}\,\eta\,k B_0\,\delta B_A$. In magnetically dominated environments relevant for magnetars, $v_A\simeq c$ and $\omega^2=k^2\ll k^2+m_a^2$, so that $a_k
\simeq
-\,g_{a\gamma}\,\eta k B_0 ~\delta B_A/(k^2+m_a^2)
$, and thus the axion energy density may be given as
\begin{equation}
\mathcal{E}_a
=
\frac12 (k^2+m_a^2)|a_k|^2
=
\frac12\,g_{a\gamma}^2\,\eta^2\,B_0^2
\frac{k^2}{k^2+m_a^2}
|\delta B_A|^2 .
\label{eq:axion_energy_density}
\end{equation}
Since Alfv\'{e}n waves are damped resistively at a rate $\Gamma_A=\eta k^2/2$, the axion production rate per unit volume reads
\begin{equation}
\mathcal{\dot E}_a =
\frac14\,g_{a\gamma}^2\,\eta^3\,B_0^2
\frac{k^4}{k^2+m_a^2}
|\delta B_A|^2 .
\label{eq:axion_production_rate}
\end{equation}
The reconnecting region is described as a current sheet of length $L$, width $W$, and thickness $\delta$, with volume $V_{\rm rec}=LW\delta$. Global simulations indicate that reconnection in magnetar magnetospheres excites
Alfv\'{e}nic disturbances on macroscopic scales comparable to the system size,
$k\sim L^{-1}$, while the current--sheet thickness only controls the validity of the
resistive closure
\cite{ThompsonDuncan2001,Lyutikov2003,LiZrakeBeloborodov2019}.
Writing $|\delta B_A|^2=\epsilon_A B_0^2$, where $ \epsilon_A \simeq
B_0^2/(B_0^2 + B_{\rm rec}^2)
\min\!\left(1,\eta L/v_A\delta^2\right)$ denotes the fraction of magnetic energy deposited into Alfv\'{e}n waves ($B_{\rm rec}$ is the reconnecting field component), integration of
Eq.~\eqref{eq:axion_production_rate} yields the total axion power per mode,
\begin{equation}
\mathcal{P}_a
=
\frac14\,g_{a\gamma}^2\,\eta^3\,B_0^4\,\epsilon_A
\frac{k^4}{k^2+m_a^2}
V_{\rm rec}.
\label{eq:axion_power_general}
\end{equation}
\textit{Detectability and mass window of reconnection--powered axion bursts.---} We now connect this production mechanism to observable quantities, determining
both the experimental sensitivity of radio telescopes and the physical domain in axion mass for which this mechanism is applicable. At a distance $d$ from the source, the spectral flux density is related to the axion flux $\Phi_a$ as 
\begin{equation}
S_\nu
\equiv
\frac{d\Phi_a}{d\nu}
=
\frac{1}{4\pi d^2}\frac{d\mathcal{P}_a}{d\nu}.
\end{equation}
Axion emission from reconnection is narrowly peaked around the frequency $\nu_a=m_a/2\pi$. The finite duration of the reconnection event, $\tau_{\rm rec}\simeq L/v_A$, sets a lower bound on the intrinsic bandwidth of the signal through $\Delta\nu_a\sim 1/\tau_{\rm rec} \simeq c/L$. For magnetar-scale current sheets this typically yields $\Delta\nu_a\sim10^2$--$10^4~\mathrm{Hz}$, which is much narrower than standard
observing bandwidths.
Provided $\Delta\nu_{\rm ch}\gtrsim\Delta\nu_a$ and $\nu_a$ is spectrally resolved, the observed spectral flux density may be written as
\begin{equation}
S_\nu
\simeq
\frac{\mathcal{P}_a^{\rm max}}{4\pi d^2\,\Delta\nu_{\rm ch}},
\label{eq:axion_spectral_flux}
\end{equation}
with $\mathcal{P}_a^{\rm max}=\mathcal{P}_a \vert_{k=m_a}$. In the remainder, we assume efficient axion--photon mode conversion in the source magnetosphere, so that the axion spectral flux directly sets the observable radio flux \cite{Prabhu2023, TercasPRL2025}. The detectability of such a narrowband transient is governed by the radiometer equation. For a telescope with effective collecting area $A_{\rm eff}$ and system
temperature $T_{\rm sys}$, the signal--to--noise ratio is
$\mathrm{SNR}
=S_\nu A_{\rm eff} \sqrt{\Delta\nu_{\rm ch}\,t_{\rm int}}/k_B T_{\rm sys}\,
$, 
where $t_{\rm int}=\min(\tau_{\rm rec},t_{\rm dump})$ is the effective integration time.
Requiring $\mathrm{SNR}\ge\mathrm{SNR}_{\rm thr}$ yields the minimum detectable
spectral flux density,
\begin{equation}
S_{\nu,\min}
=
\frac{\mathrm{SNR}_{\rm thr}\,k_B T_{\rm sys}}
{A_{\rm eff}\sqrt{\Delta\nu_{\rm ch}\,t_{\rm int}}}.
\end{equation}
Combining this with Eq.~\eqref{eq:axion_spectral_flux}, the sensitivity to the
axion--photon coupling is
\begin{equation}
g_{a\gamma}^{\rm sens}
=
\left[
\frac{
32\pi d^2\,\mathrm{SNR}_{\rm thr}\,k_B T_{\rm sys}
}{
\eta^3 B_0^4\,\epsilon_A\,L W
\sqrt{\Delta\nu_{\rm ch}\,t_{\rm int}}
}
\right]^{1/2}
m_a^{-3/2}.
\label{eq:gagamma_sensitivity}
\end{equation}
The characteristic scaling
$g_{a\gamma}^{\rm sens}\propto m_a^{-3/2}$
follows directly from the $m_a^2$ dependence of axion production in
magnetic reconnections, the narrowband nature of the burst, and the radiometer equation governing transient detection. It is therefore worth contrasting with the scaling $g_{a\gamma} \propto m_a^{1/2}$ found 
by Prabhu for the non-dissipative conversion of fast radio burst (FRB) ~\cite{Prabhu2023}. In the latter, axions act as passive intermediaries that coherently
transport pre--existing FRB emission through the dense inner
magnetosphere, with the observable signal ultimately limited by axion--photon
mode conversion in the star wind. Here, axions are produced \emph{dissipatively} as a direct consequence of magnetic reconnection within a non--ideal magnetohydrodynamic
framework. The resulting sensitivity therefore reflects the dynamics of reconnection--driven
Alfv\'enic dissipation and axion production, and not the propagation or escape of an
underlying radio signal.
\begin{figure}[t!]
\centering
\includegraphics[width=\linewidth]{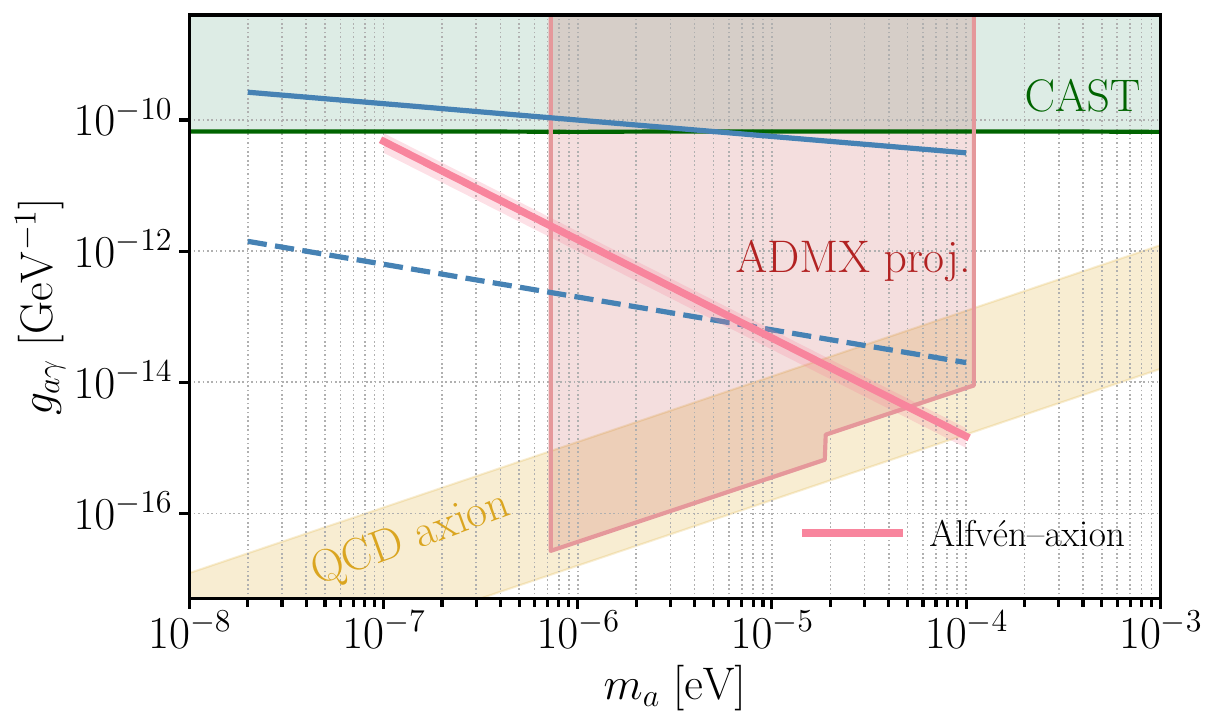}
\caption{
{\bf Sensitivity to axion--photon couplings from reconnection--driven axion bursts in
neutron star magnetospheres}. The solid curve shows the projected reach of the Alfv\'en--axion mechanism derived in
Eq.~\eqref{eq:gagamma_sensitivity}. For comparison, we depict the sensitivity estimates for axion dark-matter conversion: dashed and solid contours denote radio constraints following Ref.~\cite{Hook2018} and the plasmon dimmed situation ~\cite{TercasPRL2025}, respectively.}
\label{fig:sensitivity}
\end{figure}

The experimentally accessible axion mass window is obtained by
imposing additional constraints. At low masses, $m_a\lesssim10^{-7}\,\mathrm{eV}$, the emission frequency falls below
$\sim10~\mathrm{MHz}$, where ionospheric absorption, refractive effects, and the
rapid rise of the galactic synchrotron background severely degrade sensitivity.
In addition, when $\nu_a\ll\Delta\nu_{\rm ch}$ the signal is spectrally diluted
within a single channel, further suppressing the observed flux. At high masses, axion production is maximized for $k\sim m_a$, corresponding to a current--sheet thickness
$\delta\sim\pi/m_a$. As $m_a$ increases, this thickness eventually approaches kinetic plasma scales, such as the electron skin depth or Larmor radius, typically
$\ell_{\rm kin}\sim10^{-2}$--$1~\mathrm{m}$ in magnetar magnetospheres.
Once $\delta\lesssim\ell_{\rm kin}$, the resistive aMHD description breaks down and
the present framework no longer applies. Combining these considerations, reconnection--powered axion bursts are sensitive in
the mass range
\begin{equation}
10^{-7}\,\mathrm{eV}
\;\lesssim\;
m_a
\;\lesssim\;
10^{-4}\text{--}10^{-3}\,\mathrm{eV},
\end{equation}
In Fig.~\ref{fig:sensitivity}, we depict the projected sensitivity to the
axion--photon coupling obtained from Eq.~\eqref{eq:gagamma_sensitivity} by adopting
fiducial parameters representative of a nearby Galactic magnetar and of current
wide--field radio transient searches.
As a benchmark source we consider the magnetar SGR~1935+2154, located at a distance
$d\simeq9~\mathrm{kpc}$ and characterized by an inferred surface dipole field strength
$B_0\simeq2\times10^{14}~\mathrm{G}$
\cite{Israel2016,Borghese2020}.
The reconnecting region is modeled as a macroscopic current sheet of length and
width $L\simeq W\simeq10^4~\mathrm{m}$, consistent with magnetospheric reconnection
models and global simulations of magnetar flares
\cite{ThompsonDuncan2001,Lyutikov2003,BransgroveBeloborodov2026}.
We assume that a fraction $\epsilon_A\simeq0.1$ of the released magnetic energy is
converted into large--scale Alfv\'enic disturbances, in line with numerical studies
of relativistic reconnection and Alfv\'enic energy transport
\cite{Lyutikov2006,LiZrakeBeloborodov2019}.

We adopt an effective magnetic diffusivity
$\eta\simeq10^{-2}~\mathrm{m^2\,s^{-1}}$, and assume a relativistic Alfv\'en speed $v_A\simeq c$. For the instrumental parameters we take values characteristic of modern low-- to
mid--frequency radio arrays such as CHIME \cite{CHIME2018,Chime2025}, LOFAR \cite{vanderTol2026}, and SKA--Low \cite{Braun2015, SKALow2015}: a system temperature
$T_{\rm sys}\simeq50~\mathrm{K}$, an effective collecting area
$A_{\rm eff}\simeq10^4~\mathrm{m^2}$, and channel widths
$\Delta\nu_{\rm ch}\sim10^4~\mathrm{Hz}$
\cite{CHIME2018,LOFAR2013,SKALow2015}.
The effective integration time is set by the reconnection timescale,
$t_{\rm int}\simeq\tau_{\rm rec}=L/c$.
The sensitivity curve shown corresponds to a detection threshold
$\mathrm{SNR}_{\rm thr}=5$, typical of transient radio burst searches
\cite{Lorimer2007,CHIMEFRB2020}.

\textit{Conclusions.---} In this Letter we have developed axion magnetohydrodynamics from first principles,
retaining axion inertia and the essential physics of non--ideal plasmas.
Starting from a coupled axion--electromagnetic--plasma system, we performed a
controlled reduction to an axion--modified MHD framework and identified the regimes
in which magnetic dissipation can act as a direct and efficient source of axion
radiation.
Within this formulation, departures from ideal flux freezing acquire a clear and
unifying physical meaning: they become localized sites of dynamical axion
production.

A central outcome of our analysis is that magnetic reconnection naturally excites
Alfv\'enic and magnetosonic disturbances that hybridise with the axion field,
forming mixed axion--plasma polaritons.
These modes mediate coherent energy exchange between magnetic fields and axions,
governed by non--ideal electric fields that necessarily arise in dissipative plasma
dynamics.
In magnetically dominated environments this mechanism generically produces
burst--like axion emission accompanying reconnection--driven energy release.

Applying this framework to neutron stars and magnetars, we have shown that
reconnection events in strongly magnetized magnetospheres constitute a robust and
predictive astrophysical channel for axion production.
The resulting axion bursts are intrinsically transient, narrowband, and correlated
with magnetic activity, sharply distinguishing them from axion search strategies
based on static or adiabatic magnetic configurations.
When combined with the radiometer equation, these properties lead to a characteristic
sensitivity scaling \(g_{a\gamma}\propto m_a^{-3/2}\), within a mass window that
emerges naturally from the joint requirements of plasma physics and radio
instrumentation.

More broadly, the results presented here elevate magnetic reconnection from a purely
plasma--physical process to a potential mechanism for particle production in extreme
astrophysical environments.
Axion magnetohydrodynamics provides a coherent framework in which plasma dissipation,
astrophysical transients, and physics beyond the Standard Model are treated on the
same footing.
This perspective opens several avenues for future work, ranging from kinetic and
numerical extensions of reconnection models that incorporate axion dynamics
self--consistently, to targeted observational searches for narrowband transient
signals coincident with magnetar activity using existing and next--generation radio
facilities.
Taken together, our results establish axion magnetohydrodynamics as a fertile
interface between plasma physics, high--energy astrophysics, and fundamental
particle physics.

\bibliographystyle{apsrev4-2}
\bibliography{refs.bib}
\end{document}